\documentclass[conference]{IEEEtran}
\IEEEoverridecommandlockouts
\usepackage{cite}
\usepackage{amsmath,amssymb,amsfonts}
\usepackage{algorithmic}
\usepackage{graphicx}
\usepackage{textcomp}
\usepackage{xcolor}
\usepackage{booktabs}
\usepackage{tabu} 
\usepackage{multirow}
\usepackage{caption}
\usepackage{soul}
\usepackage{diagbox}
\usepackage{adjustbox}

\def\BibTeX{{\rm B\kern-.05em{\sc i\kern-.025em b}\kern-.08em
    T\kern-.1667em\lower.7ex\hbox{E}\kern-.125emX}}
    
\begin{document}

\title{Deep Learning-Based Visual Fatigue Detection Using Eye Gaze Patterns in VR}



\author{
    \IEEEauthorblockN{Numan Zafar}
    \IEEEauthorblockA{Dept. of Computer Science\\
    Clarkson University\\ 
    Potsdam, New York, USA\\
    \texttt{zafarn@clarkson.edu}}
    \and
    \IEEEauthorblockN{Johnathan Locke}
    \IEEEauthorblockA{David D. Reh School of Business\\
    Clarkson University\\ 
    Potsdam, New York, USA\\
    \texttt{lockejn@clarkson.edu}}
    \and
    \IEEEauthorblockN{Shafique Ahmad Chaudhry}
    \IEEEauthorblockA{David D. Reh School of Business\\ 
    Clarkson University\\ 
    Potsdam, New York, USA\\
    \texttt{schaudhr@clarkson.edu}}
}

\maketitle

\begin{abstract}
Prolonged exposure to virtual reality (VR) systems leads to visual fatigue, impairs user comfort, performance, and safety, particularly in high-stakes or long-duration applications. Existing fatigue detection approaches rely on subjective questionnaires or intrusive physiological signals, such as EEG, heart rate, or eye-blink count, which limit their scalability and real-time applicability. This paper introduces a deep learning-based study for detecting visual fatigue using continuous eye-gaze trajectories recorded in VR. We use the GazeBaseVR dataset comprising binocular eye-tracking data from 407 participants across five immersive tasks, extract cyclopean eye-gaze angles, and evaluate six deep classifiers. Our results demonstrate that EKYT achieves up to 94\% accuracy, particularly in tasks demanding high visual attention, such as video viewing and text reading. We further analyze gaze variance and subjective fatigue measures, indicating significant behavioral differences between fatigued and non-fatigued conditions. These findings establish eye-gaze dynamics as a reliable and nonintrusive modality for continuous fatigue detection in immersive VR, offering practical implications for adaptive human-computer interactions.

\end{abstract}


\section{Introduction}

Virtual reality (VR) has evolved rapidly across various domains, including education~\cite{gao2021digital, minnekanti2024classesinvr}, emergency simulation~\cite{de2020use, rettinger2022defuse}, healthcare~\cite{liu2024facilitating}, gaming~\cite{vatsal2024analysis}, and finance~\cite{mathis2022can}. VR is being employed in therapeutic applications, such as treating anxiety disorders~\cite{chard2022virtual}, phobias~\cite{ramalhoto2024phobeevr}, physical rehabilitation~\cite{o2024measuring, amin2024effectiveness}, and fitness~\cite{hayes2024more}. 
VR devices are becoming increasingly affordable and portable, leading to an expected increase in their adoption for daily use~\cite{li2024evaluating}. However, prolonged exposure to VR may cause users to experience visual fatigue~\cite{hua2017enabling}, which can affect concentration, task accuracy, and comfort. Detecting fatigue in VR ensures user engagement, safety, and performance, particularly in high-stakes or prolonged-use scenarios such as training simulations, rehabilitation programs, and immersive learning environments. Early studies rely on self-reported questionnaires~\cite{erickson2020effects} or physiological signals, including electroencephalography (EEG)~\cite{wang2025investigating}, heart rate~\cite{du2020vision}, or eye-blink patterns~\cite{souchet2022measuring}. Although informative, these approaches are often subjective, intrusive, and inadequate for continuous real-time monitoring during natural interactions. To address these limitations, eye-gaze~\cite{hu2020dgaze} emerges as a promising alternative. Integrated into modern VR headsets, eye tracking offers a passive and unobtrusive means of capturing user attention and cognitive states. Features, such as fixation duration, blink rate, and saccadic movements, are associated with fatigue-related changes, indicating the potential for early and continuous fatigue detection.

In this study, we present a novel approach for automating the detection of visual fatigue in VR using deep learning classifiers. We utilize the publicly available GazeBaseVR dataset~\cite{lohr2023gazebasevr}, a large-scale longitudinal eye-tracking resource collected from 407 participants, using a VR headset across five distinct tasks. The dataset provides high-resolution binocular gaze data recorded at 250 Hz in immersive environments that are suitable for fatigue detection research. Our analysis focuses on the horizontal and vertical gaze angles (x- and y-components) of the cyclopean eye to model visual fatigue across various task conditions. We model gaze trajectories to non-intrusively detect visual fatigue. Unlike prior work, which relies on subjective surveys or physiological signals, our approach employs passive eye tracking to capture subtle variations in gaze behavior associated with fatigue. We assess state-of-the-art deep-learning classifiers on task-specific gaze segments and explore how fatigue signatures differ across various visual demands and temporal windows. Our contribution of this paper are:
\begin{enumerate}
    \item A deep learning-based framework for visual fatigue detection using cyclopean gaze trajectories, a unique binocular analysis of fatigue-related eye movement patterns.
    \item A correlation between gaze-based features and subjective fatigue ratings.
\end{enumerate}
We achieve an accuracy of up to 94\% for tasks requiring high visual attention, thereby demonstrating the efficacy of gaze-based fatigue detection. This study lays the groundwork for the development of adaptive fatigue-aware VR systems that enhance user well-being and performance.

\section{Related Work}

\begin{table*}[ht]
\captionsetup{justification=raggedright,singlelinecheck=false, font=footnotesize}
\centering
\resizebox{\textwidth}{!}{%
\begin{tabular}{lcccclc}
\hline
\textbf{Dataset} & \textbf{User} & \textbf{Session time} & \textbf{Task} & \textbf{Features} & \textbf{Study} & \textbf{Acc} \\
\hline
Fan et al.~\cite{fan2023eye}  & 33 & 30 min (2 modes) & VR gaming & Pupil Labs Eye Tracker, Blink rate & Questionnaire & - \\
\hline
Chiossi et al.~\cite{chiossi2024mind}  & 20 & $~$60 min & VR visual blur task & EEG & Questionaire (SSQ) & - \\
\hline
Sun et al.~\cite{sun2024exploration} & 19 & Multiple Short Tasks & Flicker, reading and visual search & Eye Tracking, Pupil and Blink rate & DL & 87.9\% \\
\hline
Ji et al.~\cite{ji2025evaluation}  & 40 & 20 - 30 min & VR gaming & Eye Tracking and camera recording & DL & 91.8\% \\
\hline
Wang et al.~\cite{wang2019assessment}  & 105 & 35 min & HMD Viewing & Eye Tracking and Optometric features & SVM & 90.8\% \\
\hline
\textbf{GazeBaseVR (ours)} & \textbf{407 }& \textbf{10 - 15 min} & \textbf{Five Different Tasks} & \textbf{Eye Tracking} & \textbf{DL} & \textbf{94\%} \\
\hline
\end{tabular}
}
\caption{Summary of related work, EEG stands for Electroencephalography, DL represents Deep Learning, Acc represents Accuracy}
\label{relatedwork}
\end{table*}

Recent studies have focused on visual fatigue in VR, where prolonged exposure has affected user performance, comfort, and overall experience. Hua et al.~\cite{hua2017enabling} conducted a study using subjective questionnaires and determined that the use of VR can lead to visual discomfort symptoms such as eyestrain, dizziness, and several visual fatigue. Iskander et al.~\cite{iskander2019using} investigated the effects of vergence–accommodation conflict (VAC) on eye vergence behavior, which can lead to visual fatigue. Several studies have assessed visual fatigue in VR using subjective questionnaires to evaluate the users' perceived discomfort. For instance, Fan et al.~\cite{fan2023eye} presented a study that different modes of interaction significantly influence fatigue levels and eye movement behavior. Li et al.~\cite{li2024towards} used the Computer Vision Syndrome Questionnaire (CVS-Q) to measure visual fatigue during immersive reading tasks and explored gaze redirection as a mitigation strategy. Similarly, Hirzle et al.~\cite{hirzle2022understanding} examined digital eye strain in head-mounted displays using self-reported survey data with objective eye-tracking analysis. These studies underscored the prevalent use of questionnaire-based assessments; however, their subjective nature and limited temporal resolution pose challenges for real-time fatigue monitoring in dynamic VR environments.

Previous studies have explored the use of physiological signals such as electroencephalography (EEG) and electrooculography (EOG) for the detection of visual fatigue in VR environments. Chiossi et al.~\cite{chiossi2024mind} analyzed event-related potentials (ERPs) derived from EEG recordings to identify objective markers of visual strain during the use of head-mounted display. Similarly, the AttentivU system, introduced by Kosmyna et al.~\cite{kosmyna2019attentivu}, incorporated EEG and EOG sensors into a wearable glasses form factor, facilitating the continuous monitoring of attentional states and fatigue. Zao et al.~\cite{zao2017augmenting} proposed a neuromonitoring VR headset that combines EEG and EOG to track users’ physiological responses in real time, offering potential for the development of adaptive fatigue-aware VR systems. 

Some studies have utilized image-based and eye-tracking data to detect visual fatigue. Sun et al.~\cite{sun2024exploration} presented a deep learning framework for fatigue detection using eye-tracking signals and external video recordings, while Ji et al.~\cite{ji2025evaluation} proposed an eye-movement detection algorithm using image differencing, which is sensitive to motion noise and lighting, affecting robustness in uncontrolled environments. Similarly, Wang et al.~\cite{wang2019assessment} developed two assessment models to estimate eye fatigue based on eye tracking features and focused on blink parameters and fixation behavior. As shown in Table~\ref{relatedwork}, most existing VR studies are based on datasets with a relatively small number of users and often focus on a narrow set of features (such as blinks) to infer fatigue. Blink-based indicators can be influenced by external factors, such as ambient lighting, user variability, or transient visual stimuli, which may limit their reliability for robust fatigue detection. In contrast, our approach utilizes continuous eye-movement trajectories to model fatigue-related changes more directly and reliably across various task conditions.

\section{Dataset}

We use publicly available GazeBaseVR~\cite{lohr2023gazebasevr} datasets, a large-scale longitudinal collection of binocular eye-tracking recordings collected in VR. The dataset consists of over 5,000 recordings collected from 407 college-aged participants over 26 months using eye-tracking-enabled HTC Vive Pro headset sampling at 250 Hz. Each participant performs a standardized battery of five eye-tracking tasks: vergence, smooth pursuit, video viewing, self-paced reading, and random oblique saccades.

\paragraph{Task 1: Vergence (VRG)}
The task involves focusing on a central black sphere, approximately 1 degree of visual angle (dva) in diameter, embedded within a static noise pattern. The sphere periodically shifts between two depths to stimulate convergence and divergence eye movements, completing 12 such transitions over a course of approximately one minute.

\paragraph{Task 2: Smooth pursuit task (PUR)}
A small black dot traverses the visual field horizontally at three progressively increasing velocities (5, 10, and 20 dva/s). The participant engages in ocular tracking of this motion as the dot executes multiple sweeps from left to right, and vice versa. Each velocity phase persists for approximately 30 seconds, with brief pauses at each terminal point.

\paragraph{Task 3: Video viewing (VID)}
In the VID task, Participants watch a 38-second video clip from the animated film Big Buck Bunny, presented on a flat panel in VR. The task passively elicits natural eye movements, such as saccades and fixations, as participants follow dynamic content and characters on the screen.

\paragraph{Task 4: Reading text (TEX)}
In this task, participants read a short text of approximately 820 characters, displayed on a flat static screen. The text appears in fixed-width black font against a gray background. Following the reading task, the participants answer a straightforward multiple-choice question to promote engagement, although their responses are not included in the dataset.

\paragraph{Task 5: Random saccade task (RAN)}
A small black sphere manifests and relocates to random positions on the display every 1 to 1.5 seconds, causing rapid saccadic eye movement. The duration of the task ranges from 80 to 120 s and includes 79 target movements.
\begin{equation}
\label{eq:h}
\theta_H = \frac{180}{\pi} \, \text{atan2}\left(x, \sqrt{y^2 + z^2}\right)
\end{equation}

\begin{equation}
\label{eq:v}
\theta_V = \frac{180}{\pi} \, \text{atan2}(y, z)
\end{equation}

Each recording contains timestamp (n) measures in milliseconds and gaze direction data for individual eyes and the combine binocular gaze, referred to as the cyclopean eye. The horizontal $\theta_H$ (x) and vertical $\theta_V$ (y) gaze components of the cyclopean eye are reported in degrees of visual angle (dva). The corresponding components for the left and right eyes are denoted by lx, ly, and rx, ry, respectively. These gaze angles derive from three-dimensional gaze direction vectors using the arctangent functions in Equations~\ref{eq:h} and~\ref{eq:v}. 


\begin{table}[h]
\captionsetup{justification=raggedright,singlelinecheck=false, font=footnotesize}

    \centering

    \resizebox{\columnwidth}{!}{%
        \begin{tabular}{|l|c|c|c|}
        \hline
        \textbf{Category} & \textbf{Count} & \textbf{Age $\pm$ STD} & \textbf{HSLN $\pm$ STD} \\ 
        \hline \hline
        Number of Participants & 407 &20.96 $\pm$ 4.14 & 7.05 $\pm$ 1.48\\ 
        \hline
        Number of female Participants & 216 & 20.12 $\pm$ 2.80 & 6.93 $\pm$ 1.49\\ 
        \hline
        Number of male Participants & 188 & 21.73 $\pm$ 5.17 & 7.16 $\pm$ 1.42\\ 
        \hline
        Number of Participants with no Fatigue  & 177 & 21.18 $\pm$ 4.55 & 7.32$\pm$ 1.42\\ 
        \hline
        Number of Participants with Fatigue & 230 & 20.62 $\pm$ 3.78 & 6.83 $\pm$ 1.49 \\ 
        \hline
        \multicolumn{4}{l}{\scriptsize Note: HSLN = Hours Slept Last Night} \\
    \end{tabular}
    }
    \caption{Summary of Demographics of Dataset}
    \label{tab:demographics}
\end{table}

Participants are recruited from Texas State University and demographic information, including age, gender, and self-reported fatigue levels, is obtained through supplementary metadata. For this study, we utilize a subset of the dataset and categorize participants according to their fatigue condition. Table~\ref{tab:demographics} shows the demographic summary of the study participants. The mean age of the participants is approximately 20.96 ± 4.14 years, with a gender distribution of 216 females and 188 males. Based on self-reported fatigue levels, 230 participants experience fatigue, while 177 report no fatigue. Participants experiencing fatigue report fewer hours of sleep on average the night before the recording (6.83 ± 1.49 h) with a minimum of 1.45 hours of sleep compared to those without fatigue (7.32 ± 1.42 hours) at a minimum of 4 h, indicating a potential association between the duration of sleep and perceived fatigue.


\section{Experiment}

To detect fatigue based on eye-gaze data, we implement and evaluate six state-of-the-art deep learning models: Eye Know You Too (EKYT)~\cite{lohr2022eye}, Fully Convolutional Network (FCN)~\cite{wang2017time}, Temporal Convolutional Network (TCN)~\cite{soo2017interpretable}, InceptionTime~\cite{ismail2020inceptiontime}, Time Le-Net (TLE-Net)~\cite{le2016data}, and Multi Channel Deep Convolutional Neural Network (MCDCNN)~\cite{ismail2019deep}. We select these models for their effectiveness in time series and sequence classification tasks, which align with the nature of eye gaze behavior. Early detection is critical in immersive systems to enable timely intervention before fatigue affects user performance or safety. We segment the dataset into four different time frames (5, 10, 15, and 20 s) to assess the models' sensitivity to temporal context and evaluate performance across varying durations of gaze data. We treat each segment as an independent sample for model training and evaluation. We employ an 80-20 user split to train and test the models, ensuring each user’s data appears only in either train or test set to avoid identity leakage. This ensures that we train the models on a sufficient portion of the data while retaining sufficient unseen data to validate their generalization performance.

\paragraph{EKYT}
The EKYT~\cite{lohr2022eye} model employs a deep learning architecture for time-series classification of eye-gaze data to detect fatigue. It comprises two main components: a feature extraction backbone, SimpleDenseNet, and a classification head. The SimpleDenseNet backbone incorporates a single dense block consisting of eight 1D convolutional layers (derived from a total depth of 9), each with a kernel size of 3 and dilation rates increasing in powers of two (e.g., 1, 2, 4, etc.) up to a maximum of 64. The model sets the growth rate to 32, allowing each layer to add 32 feature maps, while dense connections enable all layers to share information effectively. The model starts with four input feature channels, which expand through a dense block. Global average pooling then reduces the temporal dimension, followed by a fully connected layer that projects the features onto a 128-dimensional embedding. The classification head includes batch normalization, ReLU activation, and applied softmax for the class output probabilities.

\paragraph{FCN}
The FCN~\cite{wang2017time} model employs a fully convolutional architecture with three 1D convolutional layers. The first layer utilizes 128 filters with a kernel size of 8, the second uses 256 filters with a kernel size of 5, and the third applies 128 filters with a kernel size of 3. Batch normalization and ReLU activation follow each convolution. After the convolutions, the model applies global average pooling across the temporal dimension and uses a softmax layer to generate class probabilities.

\paragraph{InceptionTime}
The InceptionTime~\cite{ismail2020inceptiontime} model employs a deep 1D convolutional architecture with multiple stacked inception blocks to capture multiscale temporal features. Each Inception block applies four parallel operations. Three convolutions with scaled-down kernel sizes (e.g., kernel size 2, 4, 8). One max-pooling followed by a 1×1 convolution. The block concatenates these outputs along the channel dimension, applies batch normalization, and passes them through ReLU activation. When enabled, a 1×1 bottleneck convolution reduces the number of input channels. The model stacks six blocks (depth=6) and optionally adds residual connections every three layers to improve gradient flow. After feature extraction, global average pooling reduces the sequence to a fixed-size vector. Finally, a fully connected layer outputs the logits for each class.

\paragraph{MCDCNN}
The MCDCNN~\cite{ismail2019deep} deel classifier processes each input channel independently. It uses two stacked 1D convolutional layers with 8 filters and a kernel size of 5, each followed by max pooling with a kernel size of 2. The network then flattens and concatenates the outputs from each channel to form a single feature vector. It passes this vector through a fully connected layer of 732 units. Finally, a linear layer produces class logits, and softmax activation generates the class probabilities.

\paragraph{TCN}
The TCN~\cite{soo2017interpretable} model consists of three sequential 1D convolutional layers with a kernel size of 3 and padding of 1. These layers progressively increase the number of filters from 64 to 128 and then to 256. ReLU activation follows each convolution. Global average pooling reduces the final feature map, producing a fixed-size vector regardless of input length. A softmax layer then produces the class probabilities.

\paragraph{TLE-Net}
The TLENet~\cite{le2016data} model consists of two 1D convolutional layers, followed by max pooling, and two fully connected layers. The first convolutional layer applies 5 filters with a kernel size of 5, followed by a max-pooling layer with a kernel size of 2. The subsequent convolutional layer uses 20 filters with a kernel size of 5, followed by max pooling with a kernel size of 4. The model flattens the output and passes it through a fully connected layer with 500 units, followed by a softmax layer to produce the class probabilities.

\paragraph{Loss Function}
We use the Binary Cross Entropy (BCE) loss~(\ref{eq:bceloss}) to minimize the loss and optmize the parameters of the each model.
\begin{equation}
     \mathtt{L} = (1/|T|) \Sigma_{T} BCE(pred, gt),
     \label{eq:bceloss}
\end{equation}
where $|T|$ denotes the time frame, $pred$ is the predicted label, and $gt$ represents the ground truth label.

\paragraph{Implementation Details}
We trained the model on a 12-core AMD Ryzen 9 5900X CPU at 3.7 GHz and an NVIDIA RTX 4090 GPU. All models were trained for 200 epochs. Training time ranged from 100 to 150 sec for each model. We trained a total of 120 models, derived from the combination of 5 tasks, 6 classifiers, and 4 time frames. 

\section{Results}

\subsection{Fatigue Detection using Deep Classifiers}

Tables~\ref{tab:pur}--\ref{tab:vrg} summarize the classification accuracies of each model across different VR tasks and time frames. Rows represent deep classifiers, while the columns represent the time frames in seconds. Table~\ref{tab:pur} presents the results for the PUR task, where EKYT achieves 94.5\% accuracy, followed by MCDCNN at 92\% at 10s and 20s, respectively. The predictable motion pattern of the PUR task induces consistent smooth-pursuit responses, thereby facilitating the detection of subtle fatigue-induced tracking deviations in temporal models.
\begin{table}[htb]
\centering
\begin{minipage}{0.7\columnwidth}
\captionsetup{justification=raggedright,singlelinecheck=false, font=footnotesize}
\centering
\resizebox{\textwidth}{!}{%
  \scriptsize%
  
  \begin{tabu}{c || cccc}
  	\toprule
       \diagbox{Model}{Time} & 5 & 10 & 15  & 20 \\ 
      	\hline \hline 
        EKYT        & 0.734 & 0.945 & 0.814 & 0,673  \\ 
        FCN         & 0.749 & 0.804 & 0.648 & 0.802  \\ 
        TCN         & 0.562 & 0.758 & 0.658 & 0.728  \\ 
        MCDCNN      & 0.653 & 0.779 & 0.899 & 0.919  \\ 
        TLE-NET     & 0.704 & 0.749 & 0.864 & 0.834  \\ 
        INCEPTION   & 0.583 & 0.603 & 0.719 & 0.588  \\ 
        \bottomrule
   \end{tabu}
}
\caption{PUR Classification Accuracies}
\label{tab:pur}
\end{minipage}
\end{table}

Table~\ref{tab:ran} demonstrates the accuracy of 91.5\% for the RAN task at a 15-second window using EKYT, while Table~\ref{tab:tex} indicates the peak accuracy of 91\% for the TEX task using TLE-Net and 89.5\% with FCN. 
\begin{table}[htb]
\centering
\begin{minipage}{0.7\columnwidth}
\captionsetup{justification=raggedright,singlelinecheck=false, font=footnotesize}
\centering
\resizebox{\textwidth}{!}{%
  \scriptsize%
  
  \begin{tabu}{c || cccc}
  	\toprule
      \diagbox{Model}{Time} & 5 & 10 & 15  & 20 \\ 
      	\hline \hline 
        EKYT        & 0.744 & 0.683 & 0.915 & 0.844  \\ 
        FCN         & 0.563 & 0.593 & 0.588 & 0.884  \\ 
        TCN         & 0.643 & 0.568 & 0.603 & 0.653  \\ 
        MCDCNN      & 0.553 & 0.724 & 0.905 & 0.558  \\ 
        TLE-NET     & 0.655 & 0.542 & 0.688 & 0.601  \\ 
        INCEPTION   & 0.633 & 0.588 & 0.562 & 0.583  \\ 
        \bottomrule
   \end{tabu}
}
\caption{RAN Classification Accuracies}
\label{tab:ran}
\end{minipage}
\end{table}

Table~\ref{tab:vid} presents the results for the video-viewing (VID) task. The TLE-Net and MCDCNN classifiers achieve up to 93\% accuracy. These findings demonstrate that the minimal-effort gaze patterns contain discriminative visual fatigue signals. 
\begin{table}[htb]
\centering
\begin{minipage}{0.7\columnwidth}
\captionsetup{justification=raggedright,singlelinecheck=false, font=footnotesize}
\centering
\resizebox{\textwidth}{!}{%
  \scriptsize%
  
  \begin{tabu}{c || cccc}
  	\toprule
       \diagbox{Model}{Time} & 5 & 10 & 15  & 20 \\ 
      	\hline \hline 
        EKYT        & 0.558 & 0.618 & 0.694 & 0.603  \\ 
        FCN         & 0.719 & 0.708 & 0.854 & 0.895  \\ 
        TCN         & 0.673 & 0.589 & 0.608 & 0.623  \\ 
        MCDCNN      & 0.587 & 0.788 & 0.698 & 0.904  \\ 
        TLE-NET     & 0.562 & 0.910 & 0.643 & 0.859  \\ 
        INCEPTION   & 0.683 & 0.578 & 0.603 & 0.598  \\ 
        \bottomrule
   \end{tabu}
}
\caption{TEX Classification Accuracies}
\label{tab:tex}
\end{minipage}
\end{table}
TLE-Net achieves 94\% accuracy for the VRG task, as shown in Table~\ref{tab:vrg}. The structured depth evokes vergence movements, enabling temporal models to detect fatigue-related alignment deviations.

\begin{figure*}[t]
    \centering
        \includegraphics[width=\linewidth]{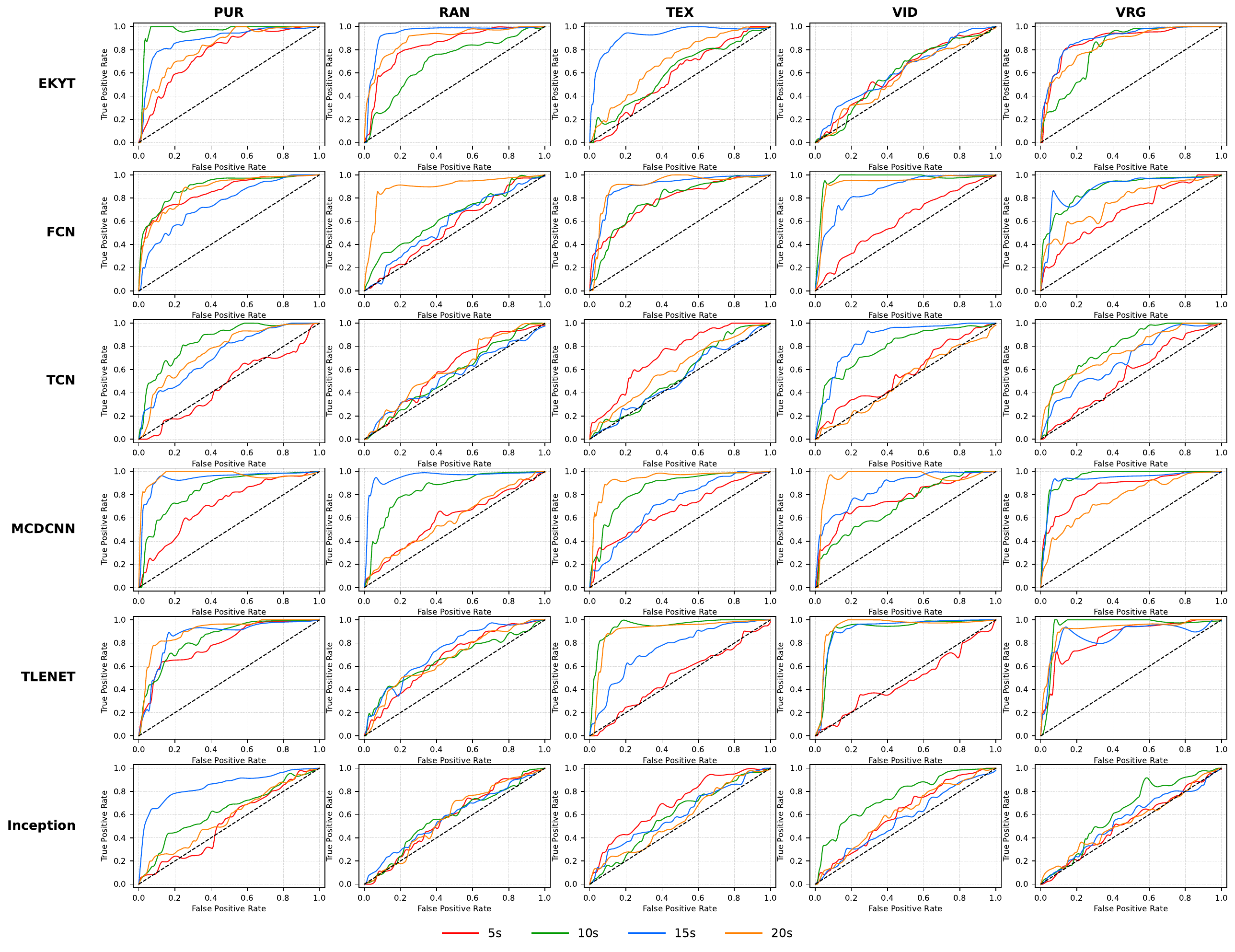}
        \caption{ROC curves for the Eye-Gaze based visual detection using Deep Classifiers.}
        \label{fig:roc}
\end{figure*}

Figure~\ref{fig:roc} shows the area under the ROC curve (AUC) for each model and task. As observe, EKYT, FCN, and MCDCNN achieve higher AUC scores across tasks, indicating their discriminative capability in visual fatigue detection. 
\begin{table}[htb]
\centering
\begin{minipage}{0.7\columnwidth}
\captionsetup{justification=raggedright,singlelinecheck=false, font=footnotesize}
\centering
\resizebox{\textwidth}{!}{%
  \scriptsize%
  
  \begin{tabu}{c || cccc}
  	\toprule
       \diagbox{Model}{Time} & 5 & 10 & 15  & 20 \\ 
      	\hline \hline 
        EKYT        & 0.598 & 0.583 & 0.608 & 0.538  \\ 
        FCN         & 0.578 & 0.929 & 0.774 & 0.915  \\ 
        TCN         & 0.573 & 0.729 & 0.824 & 0.553  \\ 
        MCDCNN      & 0.673 & 0.618 & 0.769 & 0.930  \\ 
        TLE-NET     & 0.543 & 0.914 & 0.895 & 0.935  \\ 
        INCEPTION   & 0.582 & 0.653 & 0.613 & 0.578  \\ 
        \bottomrule
   \end{tabu}
}
\caption{VID Classification Accuracies}
\label{tab:vid}
\end{minipage}
\end{table}
Our results demonstrate that deep classifiers effectively detect visual fatigue across diverse VR tasks using short segments of eye-gaze trajectories. These findings have practical implications for the design of adaptive fatigue-aware VR systems capable of fatigue monitoring.

\begin{table}[htb]
\centering
\begin{minipage}{0.7\columnwidth}
\captionsetup{justification=raggedright,singlelinecheck=false, font=footnotesize}
\centering
\resizebox{\textwidth}{!}{%
  \scriptsize%
  
  \begin{tabu}{c || cccc}
  	\toprule
       \diagbox{Model}{Time} & 5 & 10 & 15  & 20 \\ 
      	\hline \hline 
        EKYT        & 0.824 & 0.809 & 0.835 & 0.784  \\ 
        FCN         & 0.608 & 0.809 & 0.925 & 0.658  \\ 
        TCN         & 0.563 & 0.729 & 0.638 & 0.658  \\ 
        MCDCNN      & 0.788 & 0.901 & 0.915 & 0.699  \\ 
        TLE-NET     & 0.774 & 0.945 & 0.869 & 0.909  \\ 
        INCEPTION   & 0.513 & 0.603 & 0.563 & 0.577  \\ 
        \bottomrule
   \end{tabu}
}
\caption{VRG Classification Accuracies}
\label{tab:vrg}
\end{minipage}
\end{table}

\begin{figure*}[t]
    \centering
        \includegraphics[width=\linewidth]{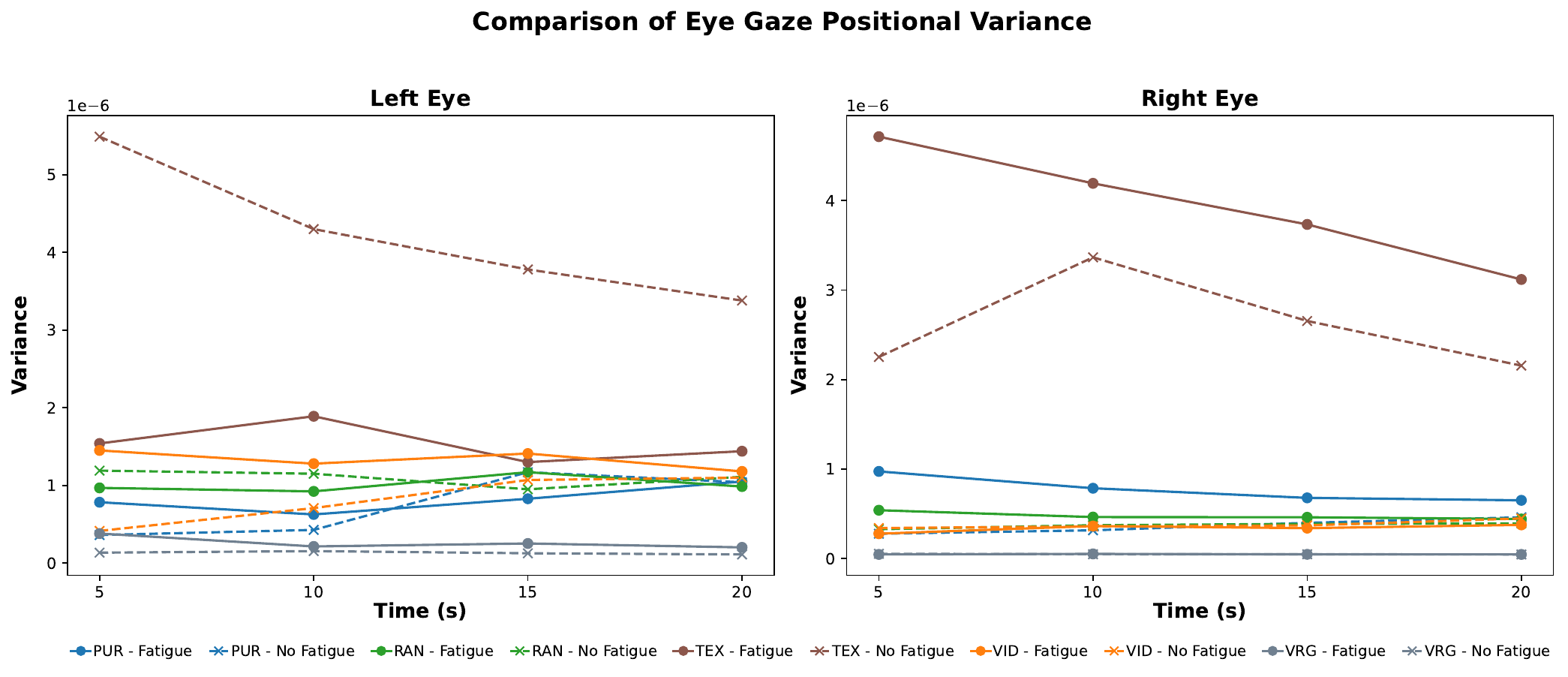}
        \caption{Eye-gaze positional variance over time across different task types under conditions of fatigue and non-fatigue.}
        \label{fig:PosVar}
\end{figure*}

\subsection{Eye-Gaze Variance Analysis}
\begin{figure*}[t]
    \centering
        \includegraphics[width=\linewidth]{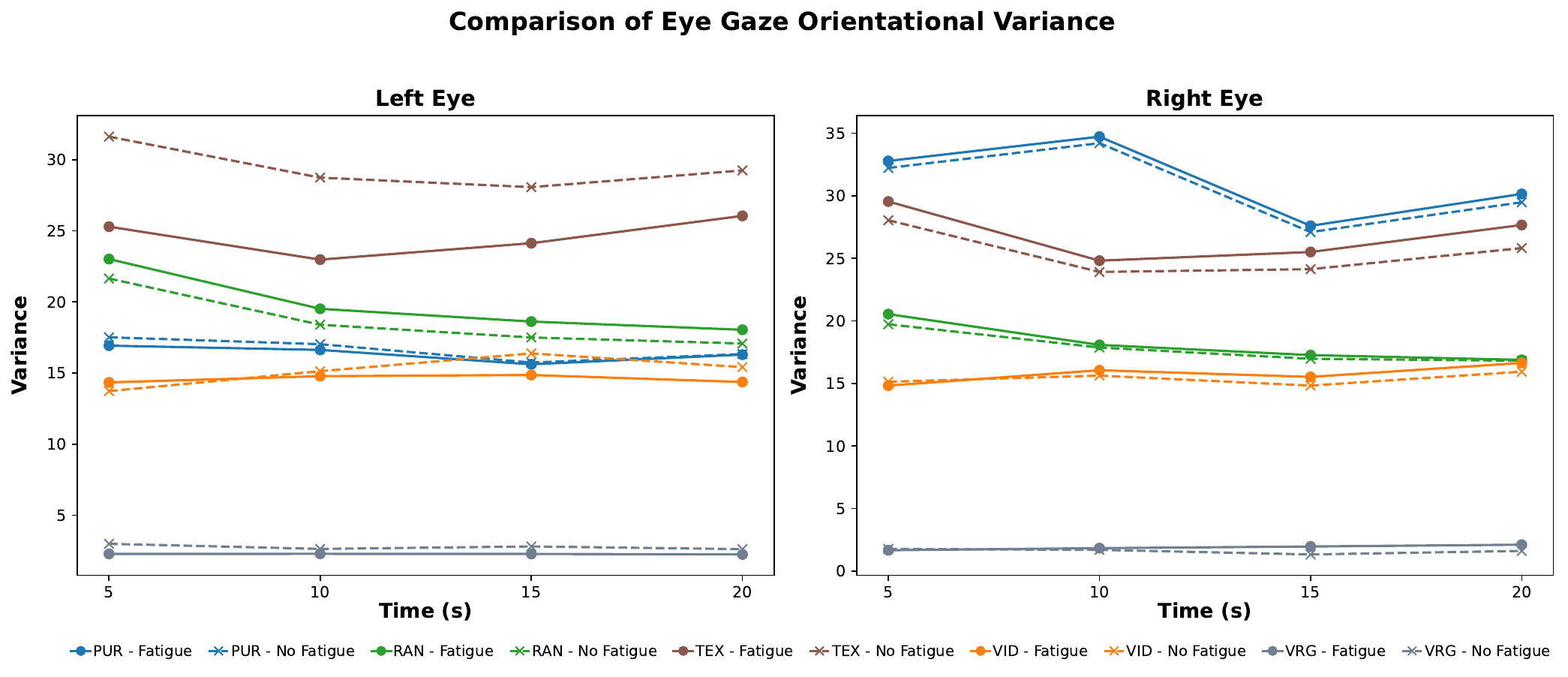}
        \caption{Variability in eye-gaze orientation over time across different task types under conditions of fatigue and non-fatigue.}
        \label{fig:RotVar}
\end{figure*}

Our analysis encompasses different aspects of user behavior and performance in a VR eye gaze-based study, including positional and rotational variance of the left and right eyes. Figure~\ref{fig:PosVar} and Figure~\ref{fig:RotVar} illustrate the differences in positional and orientational variance across five tasks for left and right eyes segmented by fatigue and no-fatigue conditions. Our analysis shows that visual fatigue is not significantly associated with short-term gaze variance in most tasks. 
\begin{itemize}
    \item In the VRG task, both left- and right-eye positional variance remain low and show no significant difference with fatigue (paired t-tests, p = 0.49 left, p = 0.78 right) while (left: p = 0.40, right: p = 0.75) orientational variance. The largely reflexive nature of vergence movements, tightly controlled by the stimulus ( alternating-depth target), likely renders them resilient to cognitive fatigue.
    \item The PUR task exhibits a slight trend of increased gaze position variance under fatigue, although it is not statistically significant (p = 0.45 left, 0.35 right) or orientational variance (left: p = 0.48, right: p = 0.52). The brief and predictable nature of the pursuit task, characterized by regular target motion lasting approximately 90 seconds, may mitigate the impact of fatigue on eye stability.
    \item Similarly, the VID (video viewing) task demonstrates virtually no difference in gaze variability between fatigue conditions (p = 0.49 left, 0.63 right) and orientational variance (left: p = 0.60, right: p = 0.75), consistent with its passive viewing nature and relatively low visual demand. The inherently guided and paced content of the video does not sufficiently challenge the oculomotor system to reveal the fatigue-induced instability.
    \item The RAN task, despite its demanding nature in terms of rapid eye movements, is not exhibit significant variance differences either (p = 0.74 left, 0.34 right) or orientational variance in the left eye (p = 0.32 or right eye (p = 0.25). We hypothesize that the inherently high variability of the RAN task, due to constantly shifting gaze, dominates any subtle changes caused by fatigue, effectively "masking" the effect. A possible reason is that the eye already moves in a highly variable, stimulus-driven manner, and there is limited scope for additional variance attributable to fatigue.
    \item In contrast to the aforementioned tasks, the TEX task demonstrates a significant impact of fatigue on gaze variance. Participants experiencing fatigue exhibit notably less stable gaze patterns during reading. We observe the positional variance of the left eye in the fatigue group is significant (p = 0.04), whereas the variance of the right eye also increases though (p = 0.20) is not statistically significant. A similar pattern is observed for orientational variance of fatigue group is significant with (p = 0.04), while right eye (p = 0.33) is not significant. This asymmetrical result may be attributed to ocular dominance or minor measurement noise differences between the eyes.
\end{itemize}

\begin{table*}[ht]
\captionsetup{justification=raggedright,singlelinecheck=false, font=footnotesize}
\centering

\resizebox{\textwidth}{!}{%
\begin{tabular}{llcccclll}
\toprule
Measure & Group & Pre-Sess & Post-Sess & $\Delta$ Mean & Pre-Group t (p) & Post-Group t (p) & Paired t (p) & $\Delta$ Group t (p) \\
\midrule
Sleepiness & No Fatigue & 1.86 & 2.16 & 0.30 & \multirow{2}{*}{$t = -6.26,\ p < 0.0001$} & \multirow{2}{*}{$t = -5.71,\ p < 0.0001$} & $t = -3.82,\ p = 0.0002$ & \multirow{2}{*}{$t = -0.15,\ p = 0.8796$} \\
           & Fatigue    & 2.43 & 2.74 & 0.31 &  &  & $t = -4.62,\ p < 0.0001$ & \\
\addlinespace
Neck Fatigue & No Fatigue & 1.39 & 1.75 & 0.36 & \multirow{2}{*}{$t = -6.21,\ p < 0.0001$} & \multirow{2}{*}{$t = -5.84,\ p < 0.0001$} & $t = -6.34,\ p < 0.0001$ & \multirow{2}{*}{$t = -0.06,\ p = 0.9485$} \\
             & Fatigue    & 1.91 & 2.27 & 0.36 &  &  & $t = -5.53,\ p < 0.0001$ & \\
\addlinespace
Physical Comfort & No Fatigue & 5.62 & 4.94 & -0.68 & \multirow{2}{*}{$t = 1.76,\ p = 0.0787$} & \multirow{2}{*}{$t = 2.07,\ p = 0.0389$} & $t = 6.05,\ p < 0.0001$ & \multirow{2}{*}{$t = 0.51,\ p = 0.6071$} \\
                 & Fatigue    & 5.39 & 4.63 & -0.75 &  &  & $t = 8.29,\ p < 0.0001$ & \\
\addlinespace
Mental Effort & No Fatigue & 1.94 & 2.01 & 0.07 & \multirow{2}{*}{$t = -4.54,\ p < 0.0001$} & \multirow{2}{*}{$t = -5.03,\ p < 0.0001$} & $t = -1.58,\ p = 0.1154$ & \multirow{2}{*}{$t = -1.19,\ p = 0.2363$} \\
              & Fatigue    & 2.36 & 2.51 & 0.14 &  &  & $t = -3.05,\ p = 0.0025$ & \\
\addlinespace
Physical Effort & No Fatigue & 1.49 & 1.54 & 0.05 & \multirow{2}{*}{$t = -2.94,\ p = 0.0035$} & \multirow{2}{*}{$t = -3.51,\ p = 0.0005$} & $t = -1.29,\ p = 0.1994$ & \multirow{2}{*}{$t = -0.85,\ p = 0.3940$} \\
                & Fatigue    & 1.74 & 1.85 & 0.11 &  &  & $t = -2.16,\ p = 0.0322$ & \\
\bottomrule
\end{tabular}
}
\caption{Comprehensive Summary of Subjective Measures Across Fatigue Groups (mean Pre vs. Post session ratings, and statistical significance of changes and group differences)}
\label{tab:subjective}
\end{table*}

\subsection{Analysis of Subjective Ratings}
For the subjective analysis, we evaluate self-reported measures of sleepiness (Stanford Sleepiness Scale 1-7), neck fatigue (1-5 scale), and physical comfort (1-6 scale) recorded before and after each session. We also assess mental and physical effort (1-5 scale) during and after each session to measure perceived changes. 
Table~\ref{tab:subjective} presents the subjective analysis and statistical comparison performed to evaluate the following:

\begin{itemize}
    \item Differences between groups before sessions (mean of pre-sess), after sessions (mean post-sess), and p values demonstrate the significance.
    \item Within-group changes (Paired t (p)) before and after the session.
    \item Differences in change scores ($\Delta$) between the two groups before and after the session.
\end{itemize}

We observe significant increases in sleepiness, neck fatigue, and physical comfort from start to end in both the groups ($p < 0.001$). However, the fatigue participants exhibit statistically significant increases in both mental effort ($p = 0.0025$) and physical effort ($p = 0.0322$), whereas these changes are not significant in non-fatigue participants. These results suggest that fatigue is associated not only with discomfort and sleepiness but also with an increased perceived cognitive and physical load~\cite{kaur2022digital, rosenfield2011computer}. Furthermore, in cross-group comparison, the fatigue group shows significantly higher subjective scores before and after the session, including sleepiness ($p < 0.0001$), neck fatigue ($p < 0.0001$), and mental effort ($p < 0.0001$). However, the change in scores ($\Delta$) is not differ significantly between the groups (e.g., $\Delta$ sleepiness: p = 0.8796), suggesting that both groups experience similar within-session changes, although the fatigue group indicates a higher mean before and after the session.

These findings demonstrate the significance of incorporating both subjective and objective metrics into fatigue detection frameworks. Eye tracking provides continuous and nonintrusive monitoring of movement, whereas subjective feedback captures the broader perceptual and physical effects of fatigue on users. A comprehensive understanding of fatigue, encompassing the visual, cognitive, and physical components, is crucial for designing adaptive VR systems to enhance user comfort, engagement, and performance.

\section{Conclusion}
This study presents a non-intrusive framework that detects visual fatigue in VR environments using deep learning models trained on cyclopean eye-gaze data. We employ the GazeBaseVR dataset to analyze five distinct VR tasks and assess six state-of-the-art time series classifiers. Among these, Eye Know You too (EKYT) and MCDCNN demonstrate performance, achieving up to 94\% and 90\% accuracy in tasks demanding a high cognitive load, such as smooth pursuit, reading, and random saccades. Our analysis of gaze variance, along with subjective ratings, further corroborates the model predictions, confirming that fatigue manifests as quantifiable changes in eye gaze behavior and perceived user discomfort. These findings underscore the potential of using continuous eye-gaze trajectories as reliable indicators of visual fatigue, particularly in tasks requiring sustained attention or inducing rapid saccadic movements. Furthermore, our results indicate that shorter gaze windows are more effective for fatigue detection.
This work lays the groundwork for developing adaptive, fatigue-aware VR systems. By integrating our fatigue detection model into VR applications, the system could proactively adjust the experience. For instance, pausing or slowing content when the user becomes fatigued, or providing alerts and guidelines to take a break. Such interventions can enhance user well being, comfort, and performance, especially in prolonged or safety-critical VR usage.

Despite the promising results, this study has limitations.
Relying on self-reported fatigue labels introduces subjectivity and may not fully capture fatigue dynamics, incorporating objective markers (e.g., blink rate or accommodative lag) alongside subjective reports could improve ground-truth reliability. The dataset's derivation from college-aged participants constrains generalizability to broader populations. Future work should focus on diverse age groups and occupations, particularly in high-risk domains such as healthcare, aviation, and remote training, where fatigue detection is critical. The current framework addresses tasks independently; there is potential for developing task-aware or task-agnostic models that generalize across VR interactions. Cross-task transfer learning should be explored to assess whether fatigue-detection models trained on one VR activity can generalize to others. Exploring self-supervised representation learning on gaze dynamics may reduce the dependence on labeled data and enhance cross-task transferability. The integration of fatigue-aware intelligence into interactive VR systems opens avenues for personalized intervention, adaptive content delivery, and cognitive workload regulation, thereby advancing the development of immersive human-centered technologies.


\bibliographystyle{IEEEtran}
\bibliography{reference} 
\end{document}